\documentclass{ws-procs9x6-cpt22}

\usepackage{bbm}

\begin{document}

\newcommand{\refeq}[1]{(\ref{#1})}
\def\etal {{\it et al.}}

\title{Massive Gravity and Lorentz Symmetry}

\author{R.\ Potting$^{1,2}$}

\address{$^1$Departamento de F\'\i sica, Faculdade de Ci\^encias e Tecnologia,\\
Universidade do Algarve, 8005-139 Faro, Portugal}

\address{$^2$Centro de Astrof\'\i sica e Gravita\c c\~ao, Instituto Superior T\'ecnico,\\
Universidade de Lisboa,  Avenida Rovisco Pais, 1049-001 Lisbon, Portugal}

\begin{abstract}
We consider Lorentz-symmetry properties of the ghost-free massive gravity theory 
proposed by de Rham, Gabadadze, and Tolley.
In particular, we present potentially observable effects in gravitational-wave propagation
and in Newton's law, including Lorentz-violating signals.
\end{abstract}

\bodymatter
\section{The de Rham--Tolley--Gabadadze action}

One of the motivations for considering massive gravity is the possibility that
modifications of General Relativity over large distances may yield a solution
to the cosmological constant problem.
A ghost-free theory of non-interacting gravitons was constructed in 1939
by Fierz and Pauli.\cite{FierzPauli}
However, attempts to generalize it to the nonlinear level failed during decades,
with the work of Boulware and Deser showing that generically such theories
will suffer from ghost instabilities.\cite{BoulwareDeser}
Recently, however, de Rham, Gabadadze, and Tolley (dRGT) showed that there exists
a non-linear extension of Fierz--Pauli massive gravity
that does not suffer from ghosts.\cite{deRhamGabadadzeTolley}
It can be shown that their model can be maximally extended to the action:%
\cite{HassanRosen}
\begin{equation}
S = \frac{1}{2\kappa}\int d^4x \sqrt{-g}
\Big( R - 2m^2 \sum_{n=0}^4 \beta_n e_n(\mathbb{X}) \Big),
\label{dRGT-action}
\end{equation}
where $R$ is the Ricci scalar, $\beta_i$ are free parameters and $\kappa = 8\pi G$.
The $4\times4$ matrix $\mathbb{X}^\mu{}_\nu$ equals $(\sqrt{g^{-1}f})^\mu{}_\nu$,
where, in addition to the usual physical metric $g_{\mu\nu}$, one assumes 
a given, nondynamical ``fiducial'' background metric $f_{\mu\nu}$.
The invariant polynomials $e_n(\mathbb{X})$
are defined through the relation
$\text{det}(\mathbbm{1}+\lambda\mathbb{X}) = \sum_{n=0}^4 \lambda^n e_n(\mathbb{X})$,
yielding $e_0(\mathbb{X})= 1$, $\> e_1(\mathbb{X})=[\mathbb{X}]$,
$\> e_2(\mathbb{X})=\frac{1}{2}([\mathbb{X}]^2 - [\mathbb{X}^2])$, \ldots.

By expressing the metric in ADM form,\cite{ArnowittDeserMisner}
using as the dynamical variables the spatial metric
together with lapse and shift variables it can be shown that the equations of
motion arising from the action \eqref{dRGT-action} yield five local degrees
of freedom, corresponding to the helicity states of a massive spin-2 particle.
Crucial in the counting is the presence of the so-called Hamiltonian constraint.

\section{Spacetime symmetries}

In this talk I will report on recent work
with Alan Kosteleck\'y\cite{KosteleckyPotting} in which
we studied the role of Lorentz symmetry in dRGT massive gravity.
Defining the vierbein $e_\mu^a$ as usual through
$g_{\mu\nu}=\eta_{ab}e_\mu^a e_\nu^b$, one can identify
local Lorentz transformations
as well as diffeomorphisms.
As was explained in the talk by Alan Kosteleck\'y at this conference,
suitable local Lorentz transformations and diffeomorphisms can be combined
to yield the so-called manifold Lorentz transformations, defined by
\begin{eqnarray}
x^\mu \to  (\Lambda^{-1})^\mu{}_\nu x^\nu, &\qquad&
g_{\mu\nu} \to (\Lambda^{-1})^\rho{}_\mu (\Lambda^{-1})^\sigma{}_\nu g_{\rho\sigma}, \nonumber\\
e_\mu{}^a  \to  (\Lambda^{-1})^\rho{}_\mu \Lambda^a{}_b e_\rho{}^b, &\qquad&
f_{\mu\nu} \to f_{\mu\nu},
\end{eqnarray}
for spacetime-independent Lorentz transformations $\Lambda$.
These are the analogues in approximately Minkowski spacetime of
global Lorentz transformations in Minkowski spacetime.
The dRGT action is
invariant under manifold Lorentz transformations if $f_{\mu\nu} \propto \eta_{\mu\nu}$,
but is otherwise Lorentz violating\cite{Kostelecky} due
to the presence of the background $f_{\mu\nu}$.

In Ref.\ [\refcite{KosteleckyPotting}] a careful study was done of the static solutions
of the dRGT potential, for flat fiducial metrics.
It was shown that the four-parameter potential has a highly nontrivial structure of
extrema and saddle points, depending on the values of the parameters.
Stability of these solutions was investigated by using the technique of bordered Hessians.
The surface generated by the Hamiltonian constraint and the positions of the solutions
on its connected sheets were used to establish global and absolute stability properties.
We concluded that extrema of the potential are invariant
under manifold Lorentz transformations,
while the saddle point solutions are Lorentz violating,
with maximally four broken Lorentz generators.

\section{Linearized massive gravity}

The action \eqref{dRGT-action} yields the equations of motion
\begin{equation}
G_{\mu\nu} + \frac{m^2}{2}\sum_{n=0}^3(-1)^n\beta_n
\left(g_{\mu\alpha}Y^\alpha_{(n)\nu}
+ g_{\nu\alpha}Y^\alpha_{(n)\mu}\right) = \kappa T_{\mu\nu},
\label{eom}
\end{equation}
where
$Y_{(n)}(\mathbb{X}) = \sum_{k=0}^n (-1)^k\mathbb{X}^{n-k} e_k(\mathbb{X})$.
Writing $g_{\mu\nu} = \eta_{\mu\nu} + h_{\mu\nu}$ and 
$f_{\mu\nu} = \eta_{\mu\nu} + \delta f_{\mu\nu}$
it follows that, to first order in $h_{\mu\nu}$ and $\delta f_{\mu\nu}$,
\begin{equation}
\mathbb{X} \approx 
\mathbbm{1} + \tfrac{1}{2}\eta^{-1}\delta f 
- \tfrac{1}{2}\eta^{-1}h + \tfrac{1}{8}\eta^{-1}\delta f\,\eta^{-1}h 
- \tfrac{3}{8}\eta^{-1}h\,\eta^{-1}\delta f.
\end{equation}
The equations of motion \eqref{eom} become
\begin{align}
&
G^{\rm L}_{\mu\nu} +\frac{m^2}{2}\sum_{n=0}^3\beta_n  
\biggl\{
2\binom{3}{n}\bigl(\eta_{\mu\nu} 
+ h_{\mu\nu}\bigr) + \binom{2}{n-1} \left(h_{\mu\nu} - \delta f_{\mu\nu} 
- \eta_{\mu\nu}\bigl[h - \delta f\bigr]\right) \nonumber \\
&\hskip20pt
{} + \left(\tfrac{1}{2}\binom{1}{n-2} + \binom{2}{n-1}\right)
\bigl[\delta f\bigr] h_{\mu\nu} 
+ \tfrac{1}{2}\binom{1}{n-2}\bigl[h\bigr]\delta f_{\mu\nu} \nonumber \\
&
\hskip20pt
{} - \tfrac{1}{2}\binom{1}{n-1}\bigl[\delta f\,\eta^{-1}h\bigr]\eta_{\mu\nu}
- \tfrac{1}{2}\binom{1}{n-2}
\bigr[\delta f\bigr]\,\bigl[h\bigl]\eta_{\mu\nu}\nonumber \\
&
\hskip20pt
{} - \left(\tfrac{3}{4}\binom{2}{n-1} 
- \tfrac{1}{2}\binom{1}{n-1}\right)
\bigl(h\,\eta^{-1}\,\delta f + \delta f\,\eta^{-1}\,h\bigr)_{\mu\nu}
\biggr\} = \kappa T_{\mu\nu} ,
\label{eom_1st-order}
\end{align}
where $G^{\rm L}_{\mu\nu}$ is the linearized Einstein tensor
and $[X] = \eta^{\mu\nu}X_{\mu\nu}$.
It is usual to
require that $h_{\mu\nu} = \delta f_{\mu\nu} = T_{\mu\nu} = 0$ satifies
the equations of motion~\eqref{eom_1st-order}, yielding the constraint
$\beta_0 + 3\beta_1 + 3\beta_2 + \beta_3 = 0$.
Moreover, we normalize the mass $m$ such that
$\beta_1 + 2\beta_2 + \beta_3 = 1$.

Consider now a nontrivial fiducial background metric $\delta f_{\mu\nu} \ne 0$,
it follows that spacetime now has nonzero curvature in the absence of matter!
In order to simplify the further analysis we will assume a special constant background
energy--momentum tensor
\begin{equation}
\kappa T_{\mu\nu} = 
-\frac{m^2}{2}\left(\delta f_{\mu\nu} - \eta_{\mu\nu}[\delta f]\right) .
\label{special-Tmunu}
\end{equation}
For this special choice, $h_{\mu\nu} = 0$ solves the equations of motion.

\section{Gravitational waves}

In order to investigate the propagation of gravitational waves,
we define the Fourier transform of $h_{\mu\nu}$ through
$ h_{\mu\nu}(x) = (2\pi)^{-4}\int d^4p\,e^{-ip\cdot x}\tilde h_{\mu\nu}(p)$.
For $\delta f_{\mu\nu} = 0$,
$\tilde{h}_{\mu\nu}$ satisfies the conditions
$\tilde h_{\mu\alpha} p^\alpha = 0$ and $\tilde h_\mu{}^\mu = 0$,
thus yielding $10-5=5$ propagating modes.
These can be identified with the helicity eigenstates
of a massive spin-two field,
$\tilde h_{\mu\nu}^{(n)}$, with $n = 0$, $\pm1$, $\pm2$.
All modes satisfy the massive dispersion relation
$(p^2 + m^2)\tilde h^{(n)}_{\mu\nu} = 0$.

For nonzero $\delta f_{\mu\nu}$ the situation becomes more complicated.
The equations of motion \eqref{eom_1st-order} can be cast in the form
\begin{equation}
(p^2 + m^2)\tilde h_{\mu\nu} =
\frac{c_2m^2}{2} S_{\mu\nu}{}^{\alpha\beta} \tilde h_{\alpha\beta}
\label{eom_grav-waves}
\end{equation}
where $c_2 = \sum_n \beta_n\left(\binom{1}{n-1} -\tfrac{3}{2} \binom{2}{n-1}\right)$.
The quantities $S_{\mu\nu}{}^{\alpha\beta}$ are tensor coefficients
depending on the momentum $p^\mu$ and on $\delta f_{\mu\nu}$.
It is convenient to expand them in a momentum-dependent orthonormal basis
spanned by $p^\mu$ and three other, spacelike vectors.
It is then straightforward to work out the expressions
$S_{\mu\nu}{}^{\alpha\beta}\tilde h^{(n)}_{\alpha\beta}$ for any helicity mode $n$,
which are well-defined linear combinations of the five helicity modes.

Equation~\eqref{eom_grav-waves} can be solved by constructing the eigenstates
of $S_{\mu\nu}{}^{\alpha\beta}$.
For general $\delta f$, we find a ``pentarefringence'' effect:
each of these eigenstates solves Eq.\ \eqref{eom_grav-waves} with
a (slightly) different dispersion relation.
These pentarefringence effects are momentum (and direction) dependent.
The modes can be sub- or superluminal,
a result typical of Lorentz-violating theories.
For details, see Ref.\ [\refcite{KosteleckyPotting}].

\section{Corrections to Newton's law}
  
Next we study the effects of the extra terms in the equation of motion~\eqref{eom_1st-order} on Newton's law.
Writing the momentum-space linearized modified Einstein equation as
$\tilde O_{\mu\nu}{}^{\alpha\beta}\tilde h^{\alpha\beta} = 0$,
the corresponding propagator is defined to satisfy
$\tilde D_{\mu\nu}{}^{\sigma\tau}\tilde O_{\sigma\tau}{}^{\alpha\beta}
= \delta^\alpha_{(\mu} \delta^\beta_{\nu)}$.
At first order in $\delta f$ it has the form
\begin{align}
\tilde D_{\mu\nu}{}^{\alpha\beta} &=
\frac{1}{p^2+m^2}
\biggl[\delta_{(\mu}^\alpha \delta_{\nu)}^\beta
- \tfrac{1}{3}\eta_{\mu\nu}\eta^{\alpha\beta}
+ \frac{2}{m^2} p_{(\mu} p^{(\alpha} \delta_{\nu)}^{\beta)} \nonumber\\
&\qquad\qquad\qquad{}- \frac{1}{3m^2}\left(p_\mu p_\nu\eta^{\alpha\beta} + \eta_{\mu\nu} p^\alpha p^\beta
- \frac{2}{m^2}p_\mu p_\nu p^\alpha p^\beta\right)\biggr] \nonumber\\
&\quad{}- \frac{m^2}{(p^2+m^2)^2}\left[
\rho_1\,\delta f_{(\mu}^\alpha \delta_{\nu)}^\beta
+ \rho_2\,\delta f_{\mu\nu}\eta^{\alpha\beta} 
+ \rho_4\, p_\mu p_\nu\delta f^{\alpha\beta} 
+ \ldots\right],
\label{propagator-explicit}
\end{align}
where $\rho_i = \rho_i(p^2)$ are momentum-dependent scalars
(parentheses in the indices indicate symmetrization).
For given energy--momentum tensor the linearized solution
for the metric can then be expressed as
\begin{equation}
h_{\mu\nu}(x) = 2\kappa\int\frac{d^4p}{(2\pi)^4}\,e^{-ip\cdot x}
D_{\mu\nu}{}^{\alpha\beta}\tilde T_{\alpha\beta}(p),
\label{metric-propagator-linear}
\end{equation}
where $T^{\mu\nu}(x) = (2\pi)^{-4}
\int d^4p\,e^{-ip\cdot x}\tilde T^{\mu\nu}(p)$.

Consider now the gravitational potential energy between two stationary point masses
with energy--momentum tensors
\begin{equation}
T_1^{\mu\nu}(x) = M_1\,\delta^\mu_0\delta^\nu_0 \,\delta^3(\vec{x}), \qquad
T_2^{\mu\nu}(x) = M_2\,\delta^\mu_0\delta^\nu_0 \,\delta^3(\vec{x} - \vec{r}).
\label{EM-tensors_point-masses}
\end{equation}
At linear order in $h_{\mu\nu}$ the matter Lagrangian corresponding to
a given energy--momentum tensor is
$\mathcal{L}_{\rm m} \approx -\tfrac{1}{2}h^{\mu\nu} T_{\mu\nu}$.
From this it follows that the potential energy corresponding to the energy--momentum tensors
\eqref{EM-tensors_point-masses} is given by
$
U(\vec{r}) 
= \int d^3x\,h_1^{\mu\nu}(\vec{x})T_{2,\mu\nu}(\vec{x}) 
= M_2\, h_{1,00}(\vec{r})\>.
$
Using Eqs.\ \eqref{propagator-explicit} and~\eqref{metric-propagator-linear} we obtain
\begin{align}
U(\vec{r}) &= -\frac{G M_1 M_2 e^{-mr}}{9r}\biggl[
24 - \delta f_{00}\bigl((4c_2+9)mr + 8c_2 + 8\bigr)\nonumber\\
&
\qquad\qquad\qquad
- \delta f_i{}^i\left((4c_2-9)mr + 2c_2 + 4 
+ \frac{2}{mr} + \frac{4}{m^2r^2}\right)\nonumber\\
&\qquad\qquad\qquad
{}- \frac{\delta f_{ij}x^ix^j}{r^2}
\left(2c_2mr + 2c_2 - 4 - \frac{6}{mr} - \frac{12}{m^2r^2}\right)
\biggr] \nonumber\\
&\qquad
{}+ \frac{2\kappa M_1 M_2}{9m^2}\,\delta^3(\vec{r})
\left[\delta f_{00} - \tfrac{2}{3}\delta f_i{}^i\right].
\end{align}
The exponential suppression factor $e^{-mr}$ is as expected
due to the massive graviton.
The term independent of $\delta f_{\mu\nu}$ is scaled by 4/3 relative to the
gravitational potential in General Relativity,
in concordance with the van Dam--Veltman--Zakharov discontinuity.
Moreover, note that  $U(\vec{r})$ acquires terms that generally
violate rotational invariance.

\section*{Acknowledgments}

This work was supported in part by the Portuguese Funda\c c\~ao
para a Ci\^encia e a Tecnologia 
under grants SFRH/BSAB/150324/2019 and UID/FIS/00099/2019,
and by the Indiana University Center for Spacetime Symmetries.

\end{document}